\documentclass[showpacs]{elsart}
\usepackage{psfig}

\begin{document}

\begin{frontmatter}
\title{\mbox{Variance fluctuations in nonstationary time series:} a
comparative study of music genres}

\author[artes,wes]{Heather~D. Jennings},
\author[bu]{Plamen Ch.~Ivanov}, 
\author[dfte,eng]{A. M. Martins},
\author[dfte]{P. C. da Silva} and   
\author[bu,dfte,ufal]{G. M. Viswanathan}

\address[artes]{Departamento de Artes,\\ Universidade Federal de
Alagoas,   Macei\'o--AL, 57072--970, Brazil}
\address[wes]{Department of Music, Wesleyan University,
Middletown, CT 06459, USA}
\address[bu]{Center for Polymer Studies and Department of
Physics,\\  Boston University, Boston, MA 02215, USA }
\address[dfte]{Departamento de F\'\i sica Te\'orica e Experimental, \\
Universidade Federal do Rio Grande do Norte, Natal--RN, 59072-970, Brazil}
\address[eng]{Departamento de Engenharia El\'etrica, \\ Universidade
Federal do Rio Grande do Norte, Natal--RN, 59072-970, Brazil}
\address[ufal]{Departamento de F\'\i sica, \\ Universidade Federal de
Alagoas, Macei\'o--AL, 57072--970, Brazil}

\date{\today}

\begin{abstract}
An important problem in physics concerns the analysis of audio time
series generated by transduced acoustic phenomena.  Here, we develop a
new method to quantify the scaling properties of 
the local variance of nonstationary time series.  We apply this
technique to analyze audio signals obtained from selected genres of
music. We find quantitative differences in the correlation properties
of high art music, popular music, and dance music.
We discuss the relevance of these
objective findings in relation to the subjective experience of music.

\end{abstract}


\end{frontmatter}

\section{Introduction}

An important problem in physics 
concerns the study of
sound.  Music consists of a complex Fourier superposition of
sinusoidal waveforms.  A person with very good hearing can hear
continuous single frequency (``monochromatic'') musical tones in the
range 20~Hz to 20~kHz~\cite{pierce}.  Audio CD players can
reproduce high fidelity music using a 44~kHz sampling rate for two
channels of 16 bit audio signals, corresponding to a maximum audible
frequency of 22~kHz~\cite{pierce,dsp}, according to the Nyquist
sampling theorem. In practice, band pass or other filters limit the
range of frequencies to the audible spectrum referred to above.
Systematic studies of the amazing complexity of music have focused
primarily on using FFT- or DFT-based spectral techniques that detect
power densities in frequency
intervals~\cite{pierce,1/f,voss-nature,gabor2001}.
For example, $1/f$-type noise in music has received considerable
attention~\cite{1/f}.  Another approach to musical complexity involves
studies of the entropy and of the fractal dimension of pitch
variations in music~\cite{entropy-df}.
Such systematic analyses have shown that music has interesting scaling
properties and long-range correlations.  
However, quantifying the
differences between qualitatively different categories of
music~\cite{genre,AI} still remains a challenge.  

Here, we adapt recently developed methods of statistical physics that
have found successful application in studying financial time
series~\cite{econo}, DNA sequences~\cite{bj} and heart rate
dynamics~\cite{heart}.  Specifically, we develop a new adaptation of
{\it Detrended Fluctuation Analysis} (DFA)~\cite{dfa,hu,chen} to study
nonstationary fluctuations in the local variance~\cite{econo} of time
series---rather than in the original time series---by calculating a
function $\alpha(t)$ that quantifies correlations on time scale $t$.
This method can detect deviations from uniform power law
scaling~\cite{bj,heart,hu,chen} embedded in scale invariant local
variance fluctuation patterns.  We apply this new method to study
correlations in highly nonstationary local variance (i.e., loudness)
fluctuations occurring in audio time series~\cite{voss-nature,econo}.
We then 
study the relationship of such objectively measurable correlations to
known subjective, qualitative musical aspects that characterize
selected genres of music. We show that the correlation properties of
popular music, high art music, and dance music differ quantitatively
from each other.

\section{Methods}

The loudness of music perceived by the human auditory system grows as
a monotonically increasing function of the average intensity.  One
typically measures the intensity of sound signals in dB (deci-Bells or
``decibels'')~\cite{pierce,dsp,acoustic}.  Hence, one conventionally
also measures loudness in dB, even though the {subjectively perceived}
loudness scales as a non-linear function of the
intensity~\cite{acoustic}.  The subjective perception of loudness
varies according to frequency and depends also on ear sensitivity,
which in turn can depend on age, sex, medication, etc~(see, e.g.,
Refs.~\cite{pierce,dsp,acoustic}).  For all practical purposes,
however, the objective measurement of sound intensity provides a good
means to quantify loudness.  In the remainder of this article, we use
the term ``loudness'' to refer to the instantaneous value of the
running or ``moving'' average of the intensity.

An important fact that deserves a detailed explanation concerns how
the human ear cannot perceive any variation in loudness (i.e.,
amplitude modulation) that occurs at frequencies $f > 20~$Hz.  Humans
hear frequencies in the audible range $20$~Hz~$<f<20$~kHz
and therefore 
do not perceive amplitude modulation or
instantaneous intensity fluctuations in this frequency range
as variations in loudness, but rather as having constant loudness. We
briefly explain this point as follows.  We can consider the the human
auditory system, in a limiting approximation, as a time-to-frequency
transducer that operates in the ``audible'' range of
20~Hz~$<f<$~20~kHz.  Any monochromatic signal in this frequency range
will lead to the perception of an audible tone of that same frequency
or ``pitch.''  A linear combination of such signals can give a
number of impressions to the human ear, depending on the exact Fourier
decomposition of the signal. Specifically, a combination of
monochromatic signals may sound as having a nontrivial
``timbre,''~\cite{pierce,acoustic} and if the signal frequencies have
special arithmetic relationships, then they may sound as a
``harmony''~\cite{pierce,acoustic}.  Beats and heterodyning, for two
or more closely spaced frequencies, can also arise. Most importantly,
a linear superposition of monochromatic signals can sound either as
having constant loudness, or else as having varying loudness. We
discuss this last point in some detail:

If a monochromatic carrier signal $U$ of frequency $f$ becomes
amplitude modulated by a modulating signal $v$ of frequency $f_M \ll
f$, then the Fourier decomposition of the modulated signal $Uv$ will
include monochromatic sidebands of sum and difference frequencies
$f\pm f_M$, but no power at frequency $f_M$~\cite{panter}.  Moreover,
amplitude modulation with $f_M<$~20~Hz results in sidebands close to
the carrier frequency, whereas $f_M>$~20~Hz leads to significant
changes in the perceived sound timbre, due to the distant sidebands
$f\pm f_M.$ Indeed, if $f_M>$~20~Hz, the sidebands fall far enough
away from the carrier to enable the ear to pick up the sidebands as
having distinct frequencies, thereby leading to the perception of a
changed timbre. Only if $f_M<$~20~Hz do the sidebands fall
sufficiently close to the carrier to fool the auditory system into
perceiving a monochromatic signal of varying loudness.
Specifically, humans hear $f_M<$~8~Hz as a ``tremolo'' (i.e., a
periodic oscillation in the intensity of the carrier tone), whereas
for 8~Hz~$<f_M<$~20~Hz we perceive a transition from the tremolo
effect to the timbre effect (see Refs.~\cite{pierce,acoustic} for more
information).  The reader should not confuse tremolos with vibratos,
which arise from frequency modulation rather than amplitude
modulation.

\begin{figure}
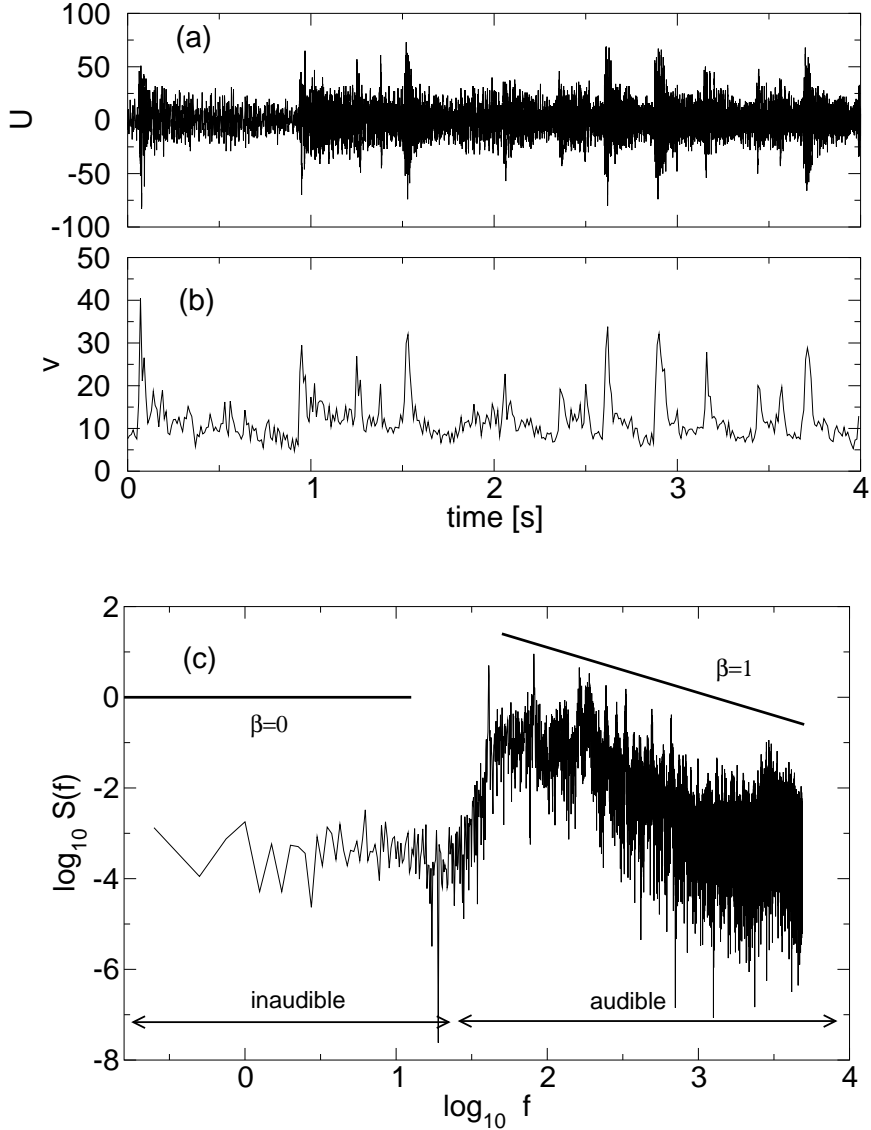


\centering

\centerline{\psfig{figure=fig1ab.eps,height=2.8truein}}

\medskip
\medskip
\medskip

\centerline{~~~\psfig{figure=fig1c.eps,height=2.8truein}}

\caption{(a)~The original signal $U(i)$ and (b)~local standard
deviation $v(j)$ for a 4~s stretch of music as a function of real time
measured in seconds.  We can relate the value of $v(j)$ to the
instantaneous loudness of the music, as described in the text.
(c)~double log plot of the power spectrum $S(f)$ as a function of
frequency $f$ measured in Hz of $U(i)$.  The human ear can only detect
monochromatic tones of frequencies in the range $20$~Hz~$<f<20$~kHz.
We instead perceive frequencies $f<20$~Hz as giving rise to melodic,
rhythmic, speech and other such structures that have time scales
$t>20^{-1}$~s. Such spectra have previously been studied
comprehensively. Note that we find 1/$f$-type behavior for audible
frequencies. The spectrum scales approximately as $S(f)\sim
f^{-\beta},$ with $\beta\approx1.$ In contrast, for lower frequencies
we find behavior more reminiscent of ``white noise,'' with
$\beta\approx 0.$ Such spectra, while useful for studying power
densities in audible frequencies, do not easily adapt to the study of
loudness fluctuations.  This forms the fundamental basis motivating
the development here of a new method that can detect deviations from
uniform power law scaling at a given time scale $t$ in the
instantaneous loudness of the music. }
\label{intro}

\end{figure}

We now devise methods suitable for studying the scaling properties of
the intensity of music signals over a range of times
scales~\cite{pierce,dsp,voss-nature}.  We begin with selected pieces
of music taken from CDs and digitize them using 8~bit sampling at
$f_s=$11~kHz. Since each piece lasts several minutes, therefore, this
``low'' 11~kHz bit sampling rate suffices for obtaining excellent
statistics.  Similarly, since we aim not to listen to music, but to
study correlations in intensity, 8~bit sound adequately satisfies
basic signal-to-noise requirements (better than $100:1$).  We choose
4~min stretches of music,
and to each piece of music assign a time series $U(i),$ where $0 \leq
U(i) \leq (2^8-1)$ and $i$ represents the sample index
(Fig.~\ref{intro}(a)).  We generate another series $v(j)$ defined as
the standard deviation of every non-overlapping 110 samples of $U(i)$.
The variance $[v(j)]^2$ thus represents the average intensity of the
sound (loudness) over intervals of $0.01$~s (Fig.~\ref{intro}(b)).
Concerning the choice of the windowing time interval, we have found
the exact value of the time interval to have little or no importance;
we have verified that our central results do not depend on the exact
value chosen, since we aim to study fluctuations in the intensity of
the signal.  We have found, e.g., that using a time interval five
times larger, $0.05$~s, equivalent to the minimum audible tone
frequency of 20~Hz, leads to no significant changes to our main
results. In this context, we note that the measurement of the loudness
of music has some similarities to the measurement of volatility in
financial markets, since in both cases the variance measurement
effectively involves a moving window of fixed but arbitrary
size~\cite{econo}.

We define
the power spectrum $S(k)$ of the signal
as the modulus
squared of the discrete Fourier transform $\tilde{U}(k)$ of $U(i)$:
\begin{equation}
S(f)\equiv |\tilde{U}(k)|^2\;\;,
\end{equation}
where $f=f_sk=11~000 \times k$ represents the frequency measured in
 Hz.  At the lowest frequencies, the spectrum appears distorted by
 artifacts of the fast Fourier transform (FFT) method.  Specifically,
 at small frequencies approaching $1/N,$ where $N$ represents the FFT
 window size, a spurious contribution arises from the treatment of the
 data as periodic with period $N$ \cite{nrc}.  The last few decades
 have seen extensive studies of the audio power spectra, considered
 nowadays well understood (Fig.~\ref{intro}(c)). The spectral power in
 the range $20$~Hz~$<~f~<20$~kHz arises due to audible sounds, while
 lower frequency contributions emerge due to the structure of the
 music on sub-audible scales larger than 20$^{-1}$~s (see
 Fig.~\ref{intro}(c)).

Since we primarily aim to study loudness fluctuations at these larger
time scales $t>20^{-1}$s, we find it more convenient to study the
power spectrum $S'(f)$ of the series $v(j)$ rather than of the series
$U(i).$ This spectrum allows us to study correlations related to
loudness at these higher time scales.  However, $v(j)$ behaves as a
highly nonstationary variable and the power spectrum of nonstationary
signals may not converge in a well behaved manner. Therefore,
conclusions drawn from such spectra may lead to questions about their
validity.  In order to circumvent these limitations, we use 
DFA.
Like the power spectrum, DFA can measure two-point correlations in
time series, however unlike power spectra, DFA also works with
nonstationary signals~\cite{bj,heart,hu,chen,henio}.

The DFA method has been systematically compared with other algorithms
for measuring fractal correlations in Ref.~\cite{taqqu2}, and
Refs.~\cite{hu,chen} contain comprehensive studies of DFA.  We use the
variant of the DFA method described in Ref.~\cite{buldyrev95}.  We
define the net displacement $y(n)$ of the sequence $v$ by \break
\mbox{$y(n)\equiv\sum_j^{n}v(j)$}, which can be thought of graphically
as a one-dimensional random walk. We divide the sequence $y(n)$ into a
number of overlapping subsequences of length $\tau,$ each shifted with
respect to the previous subsequence by a single sample. For each
subsequence, we apply linear regression to calculate an interpolated
``detrended'' walk \mbox{$y'(n)\equiv a+b(n-n_{0}).$} Then we define
the ``DFA fluctuation'' by \mbox{$F_D(\tau)\equiv\sqrt{\langle(\delta
y)^2\rangle}$,} where $\delta y\equiv y(n)-y'(n)$, and the angular
brackets denote averaging over all points $y(n)$. We use a moving
window to obtain better statistics. We define the DFA exponent
$\alpha(t)$ by
\begin{equation}
\alpha(t)
\equiv \frac{d\log{F_D(\tau)}} {d\log{(\tau+3)}} \;\;,
\label{alphaeq}
\end{equation}
where $t=100~\tau$ gives the real time scale measured in seconds.
 Uncorrelated data give rise to $\alpha=1/2,$ as expected from the
 central limit theorem, while correlated data give rise to
 $\alpha\neq1/2.$ Specifically, a value $\alpha=1/2$ corresponds to
 uncorrelated white noise, $\alpha=1$ corresponds to $1/f$-type noise
 with complex nontrivial correlations, and $\alpha=1.5$ corresponds to
 trivially correlated Brown noise (integrated white noise).
 Refs.~\cite{bj,wilson} discuss in further detail the relationship
 between DFA and the power spectrum.
A constant value of $\alpha(t)$ indicates stable
scaling~\cite{bj,heart}, while departures indicate loss of uniform
power law scaling.  
We obtain the best
statistics by studying time scales that range from $10^{-0.5}$~s to
$10$~s, hence we focus on these scales.

\section{Results}

We have recorded 10 tracks from each of 9 genres: music from the
Western European Classical Tradition (WECT), North Indian Hindustani
music, Javanese Gamelan music, Brazilian popular music, Rock and Roll,
Techno-dance music, New Age music, Jazz, and modern ``electronic''
Forr\'o dance music (with roots in traditional Forr\'o, from Northeast
Brazil).  We have chosen these genres of music somewhat arbitrarily,
noting that our main interest lies not in the music itself but rather
in developing quantitative methods of analyzing music that can---in
principle---be applied in future studies systematically to compare and
contrast diverse audio signals originating in music.

\begin{figure*}[p]
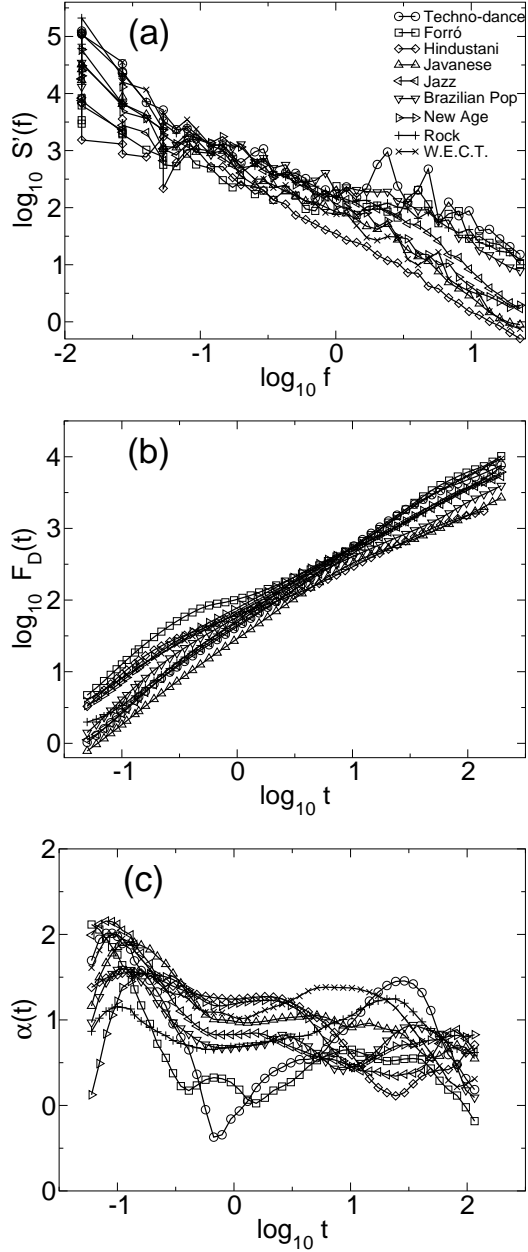


\centering

~ \psfig{figure=fig2a.eps,width=2.7truein}

\medskip

~ \psfig{figure=fig2b.eps,width=2.7truein}

\medskip

~ \psfig{figure=fig2c.eps,width=2.7truein}

\caption{(a)~Double log plot of the power spectrum $S'(f)$ of the
variable $v(j)$ for various genres of music.  For every genre we
averaged the spectrum for each individual piece of music, found using
a windows of size $2^{13}$ samples (corresponding to 81.29~s of
music), with shifts of $2^{10}$ samples (10.24s).  We applied
logarithmic binning to smooth the spectrum by averaging over windows
which grow in size as $2^{1/4}.$ These spectra suggest quantitative
differences in the scaling properties of the loudness fluctuations
that depend on the genre of music.  (b)~Double log plot of the average
DFA functions $F_D(t)$ as a function of the time scale $t$ (in
seconds) for each genre of music. (c)~Log-linear plot of the DFA
correlation exponents $\alpha(t)$ obtained from local slopes in (b),
according to Eq.~\ref{alphaeq}.  Note the striking differences between
genres, which also appear in Fig.~\ref{styles}.  }
\label{spectrum}

\end{figure*}

Fig.~\ref{spectrum}(a) shows the the power spectrum $S'(f)$ of the
series $v(j).$ As noted previously, $v(j)$ does not have stationarity
and therefore the meaning of such spectra may appear ambiguous.
Nevertheless, we can observe clear differences in the spectra of each
genre of music.  

Figs.~\ref{spectrum}(b,c) show the DFA functions $F_D(t)$ and
$\alpha(t)$, respectively. Each genre of music has a different
$\alpha(t)$ ``signature.''  In Jazz, Javanese music, New Age music,
Hindustani music and Brazilian Pop, $\alpha(t)$ decreases with $t.$
WECT music appears characterized by extremely high $\alpha(t)$ in
the region of interest from $10^{-0.5}$~s to $10^{1.0}$~s, with lower
values for rock and roll.  Techno-dance and Forr\'o music have
characteristic $\alpha(t)$ patterns marked by ``dips'' near
0.8~s. These characteristics also appear in Fig.~\ref{styles}, which
shows $\alpha(t)$ for each data set separately.

\begin{figure*}[p]

\centerline{\psfig{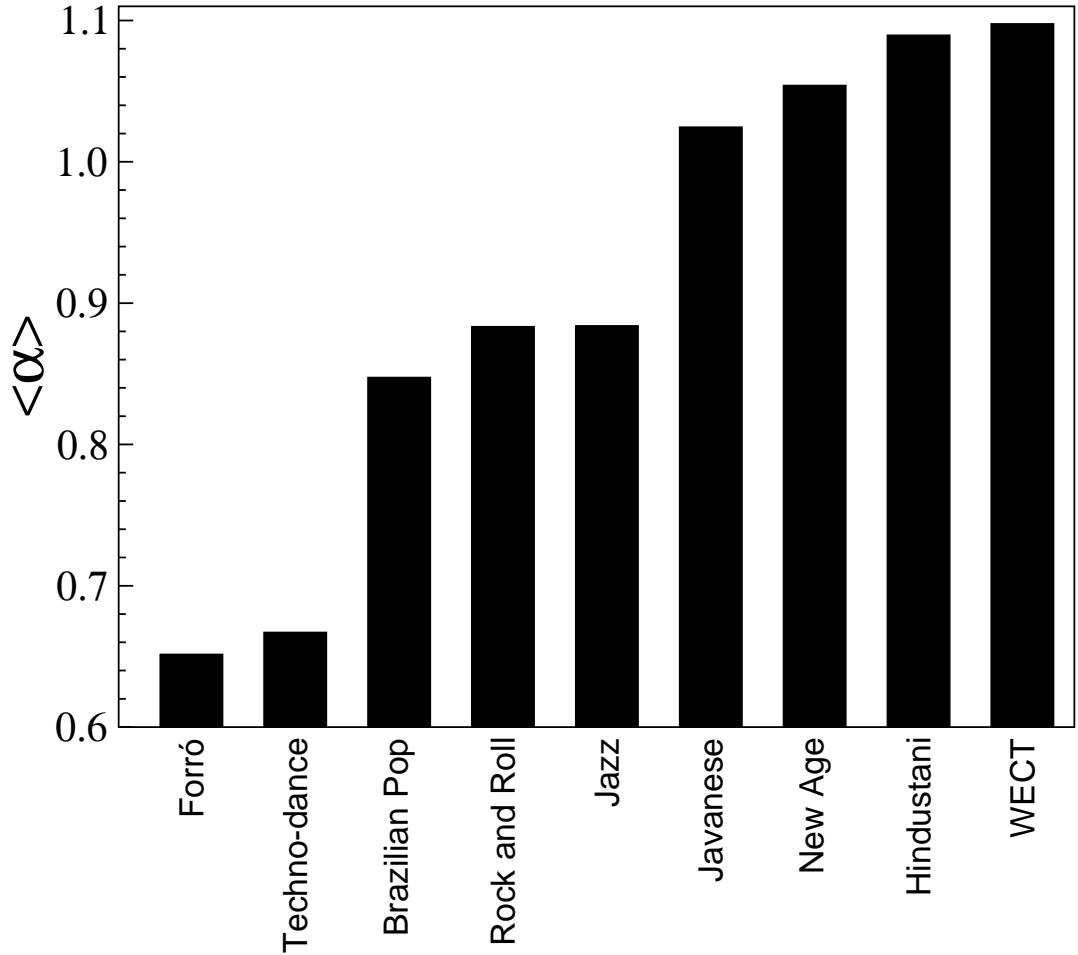}}
\bigskip

\caption{Average values $\langle\alpha\rangle$ for each genre, ranked
in increasing order.  The standard deviation of the values of $\alpha$
varies from genre to genre, but averages $\Delta \alpha =0.09$.  We
note the remarkable relationship between $\langle\alpha\rangle$ and
the music genre.  As discussed in the text, the presence of dominant
periodic trends arizing from the regular rhythmic ``beats'' can lead
to lower values of $\langle\alpha\rangle$.
The results raise the possibility that the
qualitative differences between high art, popular, and dance music
genres may be quantifiable.  }
\label{compare}
\end{figure*}

We also compute the average DFA exponent $\langle \alpha\rangle $ in
the region of interest $ 10^{-0.5}$~s$ \leq t \leq 10$~s for each
genre of music (Fig.~\ref{compare}).  We emphasize that these values
of $\alpha$ measure the scaling exponents in the variance---hence,
loudness---fluctuations of the music signals.  Any conclusions derived
from the results presented here must carefully consider this point.

\section{Discussion}

Javanese Gamelan and New Age, and to a lesser extent Hindustani and
WECT, have the values closest to $\langle\alpha\rangle=1$, corresponding to the most
complex, nontrivial correlations ($1/f$-type behavior).  We note that
WECT music has the highest value of $\langle\alpha\rangle,$ indicating that loudness
fluctuations have the strongest correlations in this genre. Hence,
from the point of view of loudness level changes, WECT music appears
the most correlated, and modern electronic Forr\'o music the least
correlated.  
None of the results reported here have a direct bearing on harmony,
melody or other aspects of music.  Our results apply only to loudness
fluctuations,
which can reflect aspects of the rhythm of the music~\cite{pierce}.

\begin{figure*}[p]
\center

 \medskip\medskip\medskip\medskip\medskip

\psfig{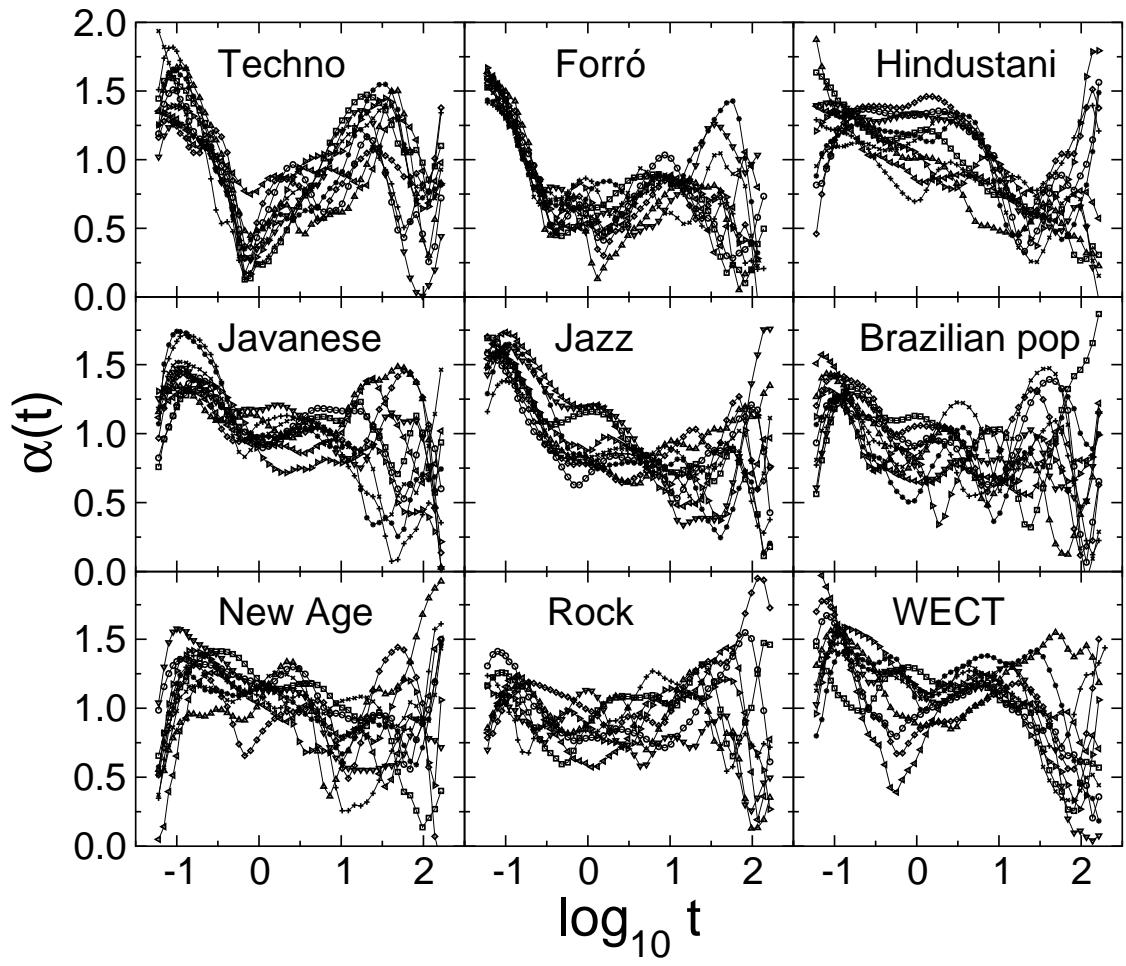}~~ ~~~~~~~~~~

 \medskip\medskip\medskip

\caption{DFA exponents $\alpha(t)$ for 9 genres of music, with 10
representative signals each. 
We have calculated $\alpha(t)$ according to Eq.~\ref{alphaeq}.  }
\label{styles}
\end{figure*}

Another observation concerns how the extremely predictable periodic
rhythmic structure of Techno-dance music and Forr\'o shows up as
minima in $\alpha(t)$ near 0.8~s (Figs.~\ref{spectrum}(c),
\ref{styles}). This finding suggests that the periodic ``beat'' of the
music, considered abstractly as a superposition of periodic trends and
the acoustic signal, leads to significant deviations from uniform
power law scaling at that time scale~\cite{bj,hu,chen}.

The above results seem to suggest that the qualitative differences
between genres---well known to music lovers---may in fact be
quantifiable.  For example, WECT music, Hindustani music and Gamelan
music, which have the highest average $\langle\alpha\rangle\approx 1$
(suggesting 
almost perfect $1/f$ scaling behavior), 
usually belong to the general category of high art music.  On the
other end, electronic Forr\'o and Techno-dance music, where periodic
tends dominate, have the lowest average $\langle\alpha\rangle,$ and
arguably belong to the category of dance or danceable music.  
The lower $\langle\alpha\rangle$ observed in these genres is due to a
a bump and horizontal shoulder in the DFA fluctuation fluctation
$F_D(t)$ that emerges at time scales corresponding to the pronounced
periodic beats~\cite{hu} (see Figs.~\ref{spectrum}(c), \ref{styles}).
Such genres 
might have evolved primarily for dancing, rather than for listening.
We can speculate  from this point of view that Jazz, Rock and Roll,
and Brazilian popular music may occupy an intermediary position
between high art music and dance music: complex enough to listen to,
but periodic and rhythmic enough to dance to.

Finally, we discuss the relevance of these findings to the possible
effects of music on the nervous system~\cite{mozart}.
Studies of heart rate dynamics using the DFA method have shown that
healthy individuals have values relatively close to
$\langle\alpha\rangle=1$, corresponding to $1/f$ correlations, while
subjects with heart disease have higher values (typically
$\langle\alpha\rangle>1.2$) that indicate a significant shift towards
less complex behavior in heart rate fluctuations, since $\alpha=1.5$
corresponds to trivially correlated Brown noise (e.g., see
\cite{heart,peng3,plamen-chaos}).  Hence, listening to certain kinds
of music may conceivably bestow benefits to the health of the
listener~\cite{mozart,depression,suicide}. The hypothesis that music with
$\langle\alpha\rangle\approx 1$ confers health benefits still requires
systematic testing.  For example, the so-called ``Mozart effect''
refers to the conjecture that listening to certain types of music may
correlate with higher test scores and more generally to
intelligence~\cite{mozart}. If ever such findings become
substantiated, then a new approach to the study of music (and perhaps
other forms of art) might become a necessity. We note, however, that
the Mozart effect has not been legitimately established as a real
phenomenon. Nevertheless, the results reported here---and more
importantly, the approach used in obtaining the results--- point
towards the possibility of
objectively analyzing subjectively experienced forms of art.
Such an approach may find relevance 
in the academic study of
music, and of art in general.

In summary, we have developed a method to study loudness fluctuations
in audio signals taken from music.  Results obtained using this method
show consistent differences between different genres of
music. Specifically, dance music and high art music appear at the
lower and upper endpoints respectively in the range of observed values
of $\langle\alpha\rangle$, with Rock and Roll, Jazz, and other genres appearing in
the middle of the range.

\section*{Acknowledgements}

We thank Ary L. Goldberger, Yongki Lee, M. G. E. da Luz, C.-K Peng, 
E.~P.~Raposo,
Luciano R. da Silva and Itamar Vidal for helpful discussions. 
We thank CNPq and FAPEAL for financial support.

\end{document}